\documentclass{aa}

\usepackage{graphicx}
\usepackage{txfonts}
\usepackage{threeparttable}
\usepackage[utf8]{inputenc}
\usepackage[T1]{fontenc}
\usepackage{amssymb}
\usepackage{amsmath}
\usepackage{hyperref}
\usepackage{xspace} 
\usepackage{ulem} 
\usepackage{orcidlink}
\usepackage{ftnxtra}
\usepackage{placeins}

\newcommand{\xmm}{{\textit{XMM-Newton}}}
\newcommand{\pn}{{\texttt{pn}}}
\newcommand{\mos}{{\texttt{MOS}}}

\begin{document}

   \title{X-ray observations of Blueberry galaxies}

   \author{B. Adamcov\'{a} \orcidlink{0009-0008-6899-4749}
          \inst{1},
          J. Svoboda \orcidlink{0000-0003-2931-0742} \inst{1},
          E. Kyritsis \orcidlink{0000-0003-1497-1134} \inst{2,3},
          K. Kouroumpatzakis \orcidlink{0000-0002-1444-2016}\inst{1},
          A. Zezas \orcidlink{0000-0001-8952-676X}\inst{2,3},
          P. G. Boorman \orcidlink{0000-0001-9379-4716}\inst{4},
          A. Borkar \orcidlink{0000-0002-9807-4520}\inst{1},
          M. B\'{i}lek \orcidlink{0000-0003-3104-3372}\inst{5,6,7},
          M. Clavel \orcidlink{0000-0003-0724-2742}\inst{8},
          P.-O. Petrucci\orcidlink{0000-0001-6061-3480}\inst{8}
          }

   \authorrunning{Adamcov\'{a} et al.}
   \institute{$^1$ Astronomical Institute of the Czech Academy of Sciences, Bo\v{c}n\'i II 1401, CZ-14100 Prague, Czech Republic\\
              \email{barbora.adamcova@asu.cas.cz}.\\
              $^2$ Physics Department, \& Institute of Theoretical and Computational Physics,   University of Crete, GR 71003, Heraklion, Greece\\
              $^3$ Institute of Astrophysics, Foundation for Research and Technology-Hellas, GR 71110 Heraklion, Greece\\
              $^4$ Cahill Center for Astronomy and Astrophysics, California Institute of Technology, Pasadena, CA 91125, USA\\
              $^5$ FZU – Institute of Physics of the Czech Academy of Sciences, Na Slovance 1999/2, Prague 182 21, Czech Republic\\
              $^6$ LERMA, Observatoire de Paris, CNRS, PSL Univ., Sorbonne Univ., 75014 Paris, France\\ 
              $^7$ Coll\`ege de France, 11 place Marcelin Berthelot, 75005 Paris, France\\
              $^8$ Univ. Grenoble Alpes, CNRS, IPAG, 38000 Grenoble, France
         }

   \date{Received ...; accepted ...}
  \abstract
   {Compact star-forming galaxies were dominant galaxy types in the early Universe. Blueberry galaxies (BBs) represent their local analogues being very compact and having intensive star formation.}
   {Motivated by high X-ray emission recently found in other analogical dwarf galaxies, called Green Peas, we probe into the X-ray properties of BBs to determine if their X-ray emission is consistent with the empirical laws for star-forming galaxies.}
   {We performed the first X-ray observations of a small sample of BBs with the {\xmm} satellite. Spectral analysis for detected sources and upper limits measured via Bayesian-based analysis for very low-count measurements were used to determine the X-ray properties of our galaxy sample.}
   {Clear detection was obtained only for 2 sources, with one source exhibiting an enhanced X-ray luminosity to the scaling relations. For the remaining 5 sources, only an upper limit was constrained, suggesting BBs to be rather underluminous as a whole. Our analysis shows that the large scatter cannot be easily explained by the stochasticity effects. While the bright source is above and inconsistent at almost the 99$\%$ confidence level, the upper limits of the two sources are below the expected distribution.
   }  
   {These results indicate that the empirical relations between the star formation rate, metallicity, and X-ray luminosity might not hold for BBs with uniquely high specific star formation rates. One possible explanation could be that the BBs may not be old enough to have a significant X-ray binary population. The high luminosity of the only bright source can be then caused by an additional X-ray source, such as a hidden active galactic nucleus or more extreme ultraluminous X-ray sources.}


   \maketitle

\begin{table*}[tbh]
\centering
\caption{Sample of Blueberry galaxies with the expected highest X-ray flux from the parent sample.}
\begin{tabular}{llcccccc}
\hline
\hline
\rule{0cm}{0.4cm}
Source & Original Name & RA (J2000) & Dec (J2000) &    $z$     &       SFR     & log(O/H)+12 & log($\mathrm{M}_*$)\\
&        & [degrees] & [degrees] &            &      [M$_\sun$/yr]  &   & [$\mathrm{M}_\sun$]    \\     
\hline
\rule{0cm}{0.3cm}
BB1 & SDSS J173501.25+570308.8\(^{(a)}\)   & 263.75512 & 57.05235 &  0.0472   &     9.74     &           8.11 & 8.6 \\ 
\rule{0cm}{0.3cm}
BB2 & 	SDSS J150934.17+373146.1\(^{(b)}\) & 227.39239 & 37.52948 & 0.03259    &     1.61     &           7.87 & 8.1  \\
\rule{0cm}{0.3cm}
BB3 & SDSS J024052.19-082827.4\(^{(a)}\)    & 40.21748 & -08.47430 & 0.0822   &     7.0     &           7.91  &  8.5     \\ 
\rule{0cm}{0.3cm}
BB4 & SDSS J085115.65+584055.0\(^{(a)}\)    & 132.81521 & 58.68195 & 0.0919   &     6.4     &           7.87  & 8.7   \\ 
\rule{0cm}{0.3cm}
BB5 & SDSS J014653.30+031922.3\(^{(b)}\)  &  26.72211 & 3.32288 & 0.04672  &         1.16   &             7.62  & 7.7    \\
\rule{0cm}{0.3cm}
BB6 & SDSS J122611.89+041536.0\(^{(a)}\)    & 186.54955 & 04.26002 & 0.0942   &     5.29     &           8.0  &  ...    \\ 
\rule{0cm}{0.3cm}
BB7 & SDSS J155624.47+480645.7\(^{(b)}\) & 239.10198 & 48.11272 & 0.05024   &       1.08     &           7.83  & 7.9   \\
\rule{0cm}{0.3cm}
BB8 & SDSS J082540.44+184617.2\(^{(b)}\) &  126.41854 & 18.77145 & 0.03792      &    0.50     &           7.79 & 7.2 \\
\hline
\end{tabular}\\
\footnotesize{\((a)\) \cite{Jaskot2019}, 
\((b)\) \cite{Yang2017}}.
\label{tab:sample}
\end{table*}

\section{Introduction}

Compact star-forming galaxies allow us to study various astrophysical processes, from star formation in extreme environments to the broader context of galaxy evolution and cosmic history. Low-mass, low-metallicity compact dwarf galaxies were abundant at the early stages of the Universe. The intense star formation ignited in such compact dwarf galaxies is proposed next to the high-energy emission from first quasars to be responsible for the reionisation of the Universe  \citep[see, e.g.][]{Robertson2010}. The main responsible mechanism of the Epoch of Reionisation, which took place after the Dark Ages in the early Universe (corresponding to the cosmological redshift $z\sim30-6$), is still hotly debated. The recently launched James Webb Space Telescope (JWST) enables the direct study of these high-redshift galaxies (see recent studies by \citealt{Harikane2023} and \citealt{Finkelstein2023}). However, a detailed study of these distant primordial galaxies is challenging due to the limited sensitivity of observations across the wavelengths. Therefore, the local galaxies with analogous properties to the early-Universe galaxies represent unique environments for multi-wavelength studies.

One of the local analogues of the high-redshift galaxies includes the Green Pea galaxies (GPs), $z\sim0.2-0.3,$ compact ($\sim$\,kpc), low-mass ($\sim10^9$\,M$_\sun$) and low-metallicity (log(O/H)+12\,$\sim$\,8.1) starburst galaxies with high star-formation rates (SFR\,$\sim10-100$\,M$_\sun$\,yr$^{-1}$), originally identified in a citizen-science Galaxy Zoo project of classification of the optical Sloan Digital Sky Survey (SDSS) observations \citep{Cardamone2009}. The characteristic green colour is due to a strong emission from ionised oxygen heated by the presence of recently formed stars. Similarly to high-redshift galaxies, GPs are also strong Ly$\alpha$ emitters \citep{Henry15, Verhamme2017, Orlitova2018}, and escape of a significant amount of ionising ultraviolet (UV) radiation (also called the Lyman continuum, LyC) have been observed in some of them \citep{Izotov2016b,Izotov2018a,Izotov2018b}. 
Most recently, \citet{Schaerer2022,Rhoads2023} revealed a high level of similarity between the JWST spectra of high-redshift galaxies and GPs. 

Somewhat surprising results came from X-ray observations of GPs with the X-ray Multi-Mirror Mission ({\xmm}). Unusually bright X-ray emission was detected in 2/3 of the sources \citep{Svoboda2019} and the reported X-ray luminosities $\sim10^{42}$\,erg\,s$^{-1}$ were a factor $>5$ larger than predicted from any theoretical or empirical relation established for star-forming galaxies (e.g. \citealt{Ranalli2003,Brorby2016}). The enhancement of the X-ray emission of the two GPs is too strong to be due to stochasticity, the enhanced population of high-mass X-ray binaries (HMXB), or ultraluminous X-ray sources (ULXs). It was also shown that the contribution from hot gas cannot explain the high X-ray luminosity \citep{Franeck2022}. One of the feasible remaining explanations is the presence of an active galactic nucleus (AGN), where the X-ray emission originates in the accretion processes onto a massive black hole \citep{Svoboda2019}. In dwarf galaxies, such massive black holes would represent a lower end of the mass range for super-massive black holes or might be considered as intermediate-mass black holes with the masses around $\sim\!\!10^4$-$10^5\,$M$_\odot$ (see studies by 
\citeauthor{Mezcua2016} \citeyear{Mezcua2016}, \citeyear{Mezcua2018}).

The GPs are by definition at redshift $z\gtrsim0.2-0.3$. \citet{Yang2017} identified a sample of local GP analogues using similar selection criteria as for the GPs (compactness, large equivalent width of optical emission lines), but limited to $z<0.1$ SDSS galaxies. Due to their lower redshift, the characteristic colour of these galaxies is blue, and therefore they were called ``Blueberries'' (BBs). Similarly, \citet{McKinney2019} and \citet{Jaskot2019} focused their sample on galaxies with the highest ionisation (high [\ion{O}{III}] / [\ion{O}{II}] line ratio), as analogues to LyC leakers. As a shorthand, we refer to both these samples as BBs. Compared to GPs, the BBs typically have lower metallicities (log(O/H)+12 $\sim$ 7.6-8.1) and lower stellar masses ($M_* < 10^8 M_\odot)$. The specific star formation rate (sSFR), defined as the star formation rate per unit of stellar mass (SFR/$M_\star$), is higher ($\log {\rm sSFR} > -8$) than for GPs. Thus, BBs not only represent good analogues to the high-redshift galaxies, but they also significantly extend the parameter space at which scaling relations between the star formation properties and X-ray luminosity can be studied.

This paper contains the first study of X-ray emission of a sample of BBs observed with the {\xmm} satellite. The paper is structured as follows: the observation sample, details of the data reduction and analysis are provided in Section~\ref{Data_reduction}, results are shown in Section~\ref{Results} and discussed in Section~\ref{Discussion}.

\begin{table*}[tbh]
\small
\centering
\caption{Observation details and results of the XSPEC and BEHR analysis.}
\begin{tabular}{llcccccccc}
\hline
\hline
\rule{0cm}{0.4cm}
Source & Obs. ID  & \multicolumn{3}{c}{Net exposure [ks]}                     & {Flux}& log($L_{\mathrm{X, Observed}}$) & log($L_{\mathrm{X, Predicted}}$) \\
 &    & {\pn}  & {\mos}1 & {\mos}2 &  [10$^{-16}$erg s$^{-1}\:$cm$^{-2}$] & [erg s$^{-1}$] & [erg s$^{-1}$]                                 \\
\hline
\rule{0cm}{0.3cm}
BB1    & 0865260101                    & 14.7                          & 21.8                           & 21.8                                       & 63$\pm 9$                & 40.5$\pm 0.3$   & 41.0                                        \\
\rule{0cm}{0.3cm}
BB2    & 0865260201                    & 38.0                          & 53.8                           & 53.8                                     & \textless{}8.1            & \textless{}39.3 & 40.4                                        \\
\rule{0cm}{0.3cm}
BB3    & 0865260301                    & 16.6                          & 23.3                           & 23.3                                      & \textless{}9.5              & \textless{}40.2 & 41.0                                        \\
\rule{0cm}{0.3cm}
BB4    & 0865260401                    & 24.7                          & 30.8                           & 30.7                                      & \textless{}2.1            & \textless{}39.7 & 41.0                                        \\
\rule{0cm}{0.3cm}
BB5    & 0865260501                    & 5.1                           & 9.9                            & 8.9                                    & \textless{}1.5            & \textless{}39.9 & 40.4                                        \\
\rule{0cm}{0.3cm}
BB7    & 0865260701                    & 23.5                          & 43.9                           & 43.9                                     & \textless{}4.5           & \textless{}39.5 & 40.2                                        \\
\rule{0cm}{0.3cm}
BB8    & 0865260801                    & 50.7                          & 59.7                           & 59.7                                       & 68$\pm 8$            & 40.4$\pm 0.2$   & 39.9                                        \\
\hline
\end{tabular}\\
\begin{tablenotes}
    \item {\textbf{Notes:}} Rest frame fluxes and luminosities (or their upper limits) in the 0.5 - 8 keV band along with the predicted luminosity values calculated using the \citet{Brorby2016} scaling relation. For the two detected sources, the fluxes were determined with the use of the XSPEC software. For the rest, the upper limits on fluxes were measured using the \texttt{BEHR} code and the \texttt{WebPIMMS} flux estimate. 
\end{tablenotes}
\label{tab:results}
\end{table*}

\begin{table*}[tbh]
\small
\centering
\caption{Parameters and results of the XSPEC fitting.}
\begin{tabular}{lccccccccc}
\hline
\hline
\rule{0cm}{0.4cm}
Source & $\mathrm{N}_\mathrm{H}$ & {C-statistics fit goodness} & {Degrees of freedom} & {Photon Index} & {Normalisation factor}               \\
\hline
\rule{0cm}{0.3cm}
BB1              & 3.20                      & 553                & 511                                          & $1.9_{-0.9}^{+1.1}$                       & $1.4 \pm 0.7$ \\
\rule{0cm}{0.3cm}
BB8              & 4.51             & 591                                     & 586                                      & $1.7_{-0.5}^{+0.6}$                   & $1.3 \pm 0.4$ \\
\hline
\end{tabular}\\
\begin{tablenotes}
    \item {\textbf{Notes:}} The Galactic $\mathrm{N}_\mathrm{H}$ is given in {$10^{20}$ cm$^{-2}$}
    and the normalisation factor is given in 10$^{-6}$ photons/keV/cm$^{2}$/s at 1\,keV.
\end{tablenotes}
\label{tab:fitresults}
\end{table*}

\section{Observation sample, data reduction and analysis}
\label{Data_reduction}

We defined the parent sample of the BBs by combining the two above-mentioned samples: 1) \citet{Yang2017} - 40 sources, and 2) \citet{Jaskot2019} - 13 sources. For all of the galaxies in the combined parent sample, we calculated the expected X-ray luminosity by using the \citet{Brorby2016} scaling relation. The parent sample of BBs includes values of SFR estimated under the assumption of \citet{Kroupa2001} initial mass function (IMF), therefore we had to modify the \citet{Brorby2016} relation, which was constrained assuming the \citet{Salpeter1955} IMF. Throughout this paper, we adopt the values based on the \citet{Kroupa2001} IMF. To use the \citet{Brorby2016} scaling relation or for comparison with different galaxy samples, we make use of the conversion described by \citet{Madau2014} (i.e. to convert SFR from \citealp{Salpeter1955} IMF to \citealp{Kroupa2001} IMF, we multiply by a constant factor of 0.67).

After using the adjusted \citet{Brorby2016} relation, we selected 8 galaxies for X-ray observations based on their expected X-ray fluxes and thus with the shortest exposure time needed for their detection in X-rays. Given that the X-ray emission of star-forming galaxies is mainly correlated with the SFR \citep{Grimm2003, Lehmer2010} and the flux naturally drops with the distance, the selected sources are those with the largest SFRs and the lowest redshift from the parent sample. Table \ref{tab:sample} lists the full names, coordinates, and the main physical properties of the selected sources. Our targets lie in the redshift range $z = 0.03 - 0.09$, their SFRs $\sim 0.5-10$ $M_\sun$\,yr$^{-1}$ (based on H$\alpha$ measurements from \citealt{Yang2017} and \citealt{Jaskot2019}), their stellar masses log~$M_\star = 7.2-8.7$ $M_\sun$ (obtained from the \textit{ugrizy} photometry for the \citealp{Yang2017} sample, and from the MPA-JHU catalogue for the \citealp{Jaskot2019} sample) and their metallicities log [O/H] + 12 $\sim 7.6-8.1$ (measured here using the prescription of \citealt{Pettini2004}).

Our selected sources except BB6 were observed by the {\xmm} satellite during January-April 2021 with the use of the EPIC cameras operating in the full-frame mode with a thin filter (for observation details see Table \ref{tab:results}). The BB6 was not observed, therefore it is excluded in the subsequent data reduction and analysis. The total exposure times of the 7 observed BBs ranged from 17 to 61\,ks per source, but after the subtraction of intervals with high background flares (see hereafter) the net exposure times shrank significantly in some cases (see Table \ref{tab:results} for the clean exposure times for each camera). The most strongly affected source was BB5, for which the useful exposure shrank from 25 to 5\,ks in the {\pn} camera (to 10 and 9\,ks in {\mos}1 and {\mos}2 cameras, respectively), and thus only a small fraction (less than 25$\%$) of the observing time could be used.

For the data reduction, we have used version 20.0.0 of the Science Analysis System software \citep[SAS;][]{gabriel2004}. First, the SAS commands \texttt{epproc} and \texttt{emproc} were used to obtain the calibrated and concatenated event lists for the {\pn} and {\mos} detectors, which were subsequently filtered using the \texttt{tabgtigen} and \texttt{evselect} tools in order not to contain intervals of high particle background. The count rate threshold of 0.4 ct s$^{-1}$ and energy range of 10 $<$ E $<$ 12 keV were considered for {\pn}, and the count rate threshold of 0.35 ct s$^{-1}$ and E $>$ 10 keV were considered for {\mos}.

The SAS script {\textsc{edetect$\_$chain}} was used to perform detection of weak sources in several energy bands (namely $0.5-10$, $0.5-1$, $1-2$, $2-4.5$ and $4.5-10$ keV) at the same time. The detection criteria are based on the {\texttt{emldetect}} tool with the minimum detection likelihood of 8. If a source was detected by this script, we followed with the spectra extraction. For a particular observation, the source and background regions were determined identically for the {\pn} and {\mos} cameras. The counts from these regions were then extracted using the \texttt{evselect} tool. The spectra of the point-like sources were extracted as circular regions around the source coordinates from SDSS with a standard radius of 30 arcsec. To define the background regions, we have used the recommendation for the {\pn} and {\mos} cameras given in the {\xmm} Calibration Technical Note XMM-SOC-CAL-TN-0018 \citep{Smith2022}. First, the background and source regions cannot overlap as out-of-time events are to be avoided. As is recommended for the {\pn} camera, the background regions were chosen to have comparable low-energy instrumental noise to the source region, that is on the same chip of the CCD detector and with a similar distance from the readout node. As the {\mos} detectors have a less severe limitation (only the same chip is required), we used the same background regions as for the {\pn} and only visually checked for any contamination by bright sources in the background regions (for the background region coordinates see Table \ref{tab:regions} in Appendix).

As the combination of spectra from the EPIC cameras is possible only if the spectra are generated with a common bin size, the common bin size of 5 eV for all EPIC cameras was chosen. For the {\pn} detector, only patterns less than 4 and the standard energy range of $0-20479$ eV were considered, for the {\mos} detector less than 12 and range of $0-11999$ eV. For the extracted spectra, the redistribution matrices were generated using the \texttt{rmfgen} task, and the ancillary files using the \texttt{arfgen}. This allowed for the combination of the spectra into EPIC combined spectrum via the \texttt{epicspeccombine} tool. For the detected sources, we used the combined spectra for the spectral analysis. For different energy bands (namely $0.5-1$, $1-2$, and $2-10$ keV), the EPIC vignetting-corrected background-subtracted images\footnote{See the guide at: https://www.cosmos.esa.int/web/xmm-newton/sas-thread-images} were created for our sources.

\begin{figure*}
    \begin{minipage}{\textwidth}
   \centering
   \begin{tabular}{cc}
   \includegraphics[width=0.5\textwidth]{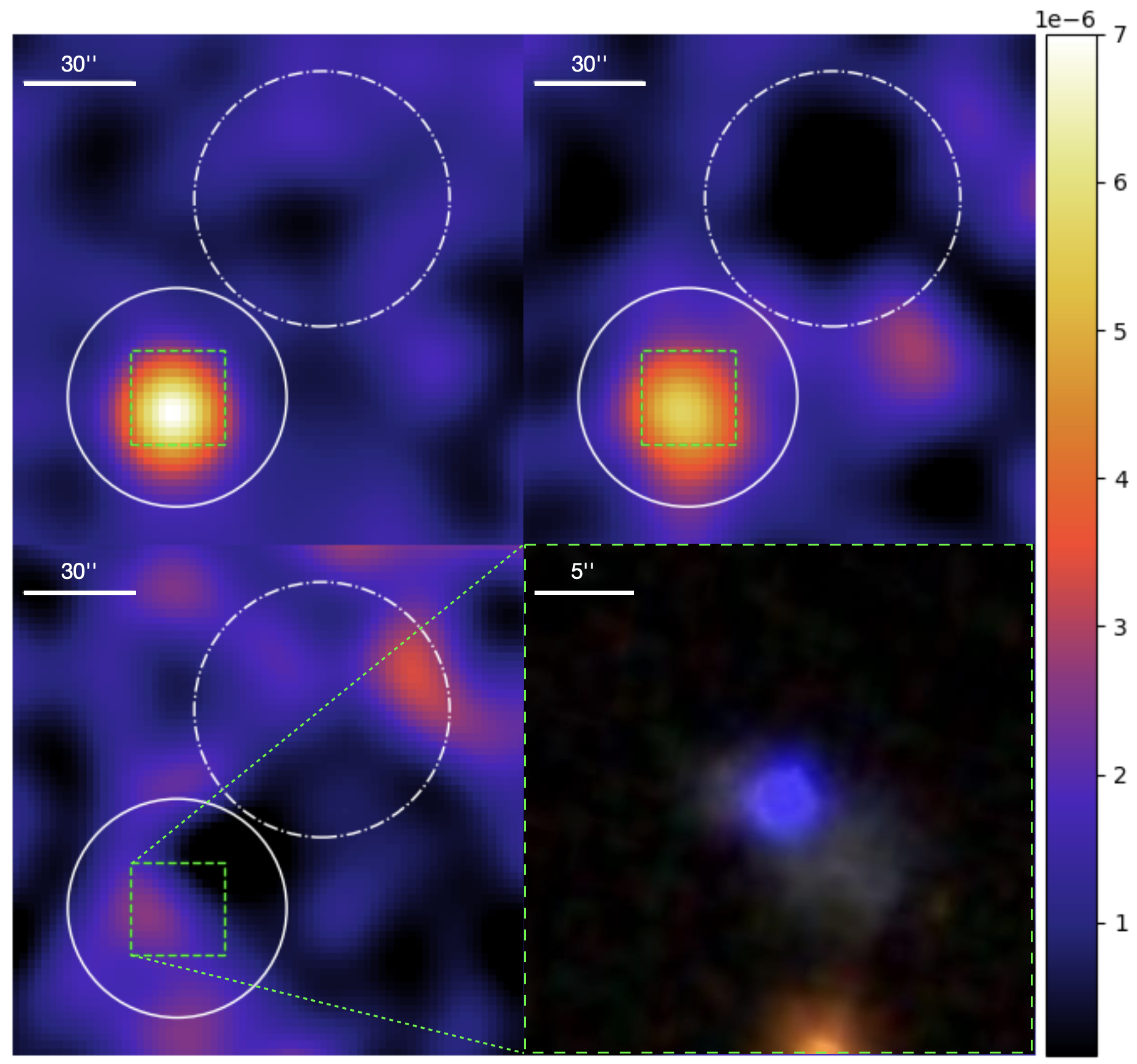}
   \includegraphics[width=0.5\textwidth]{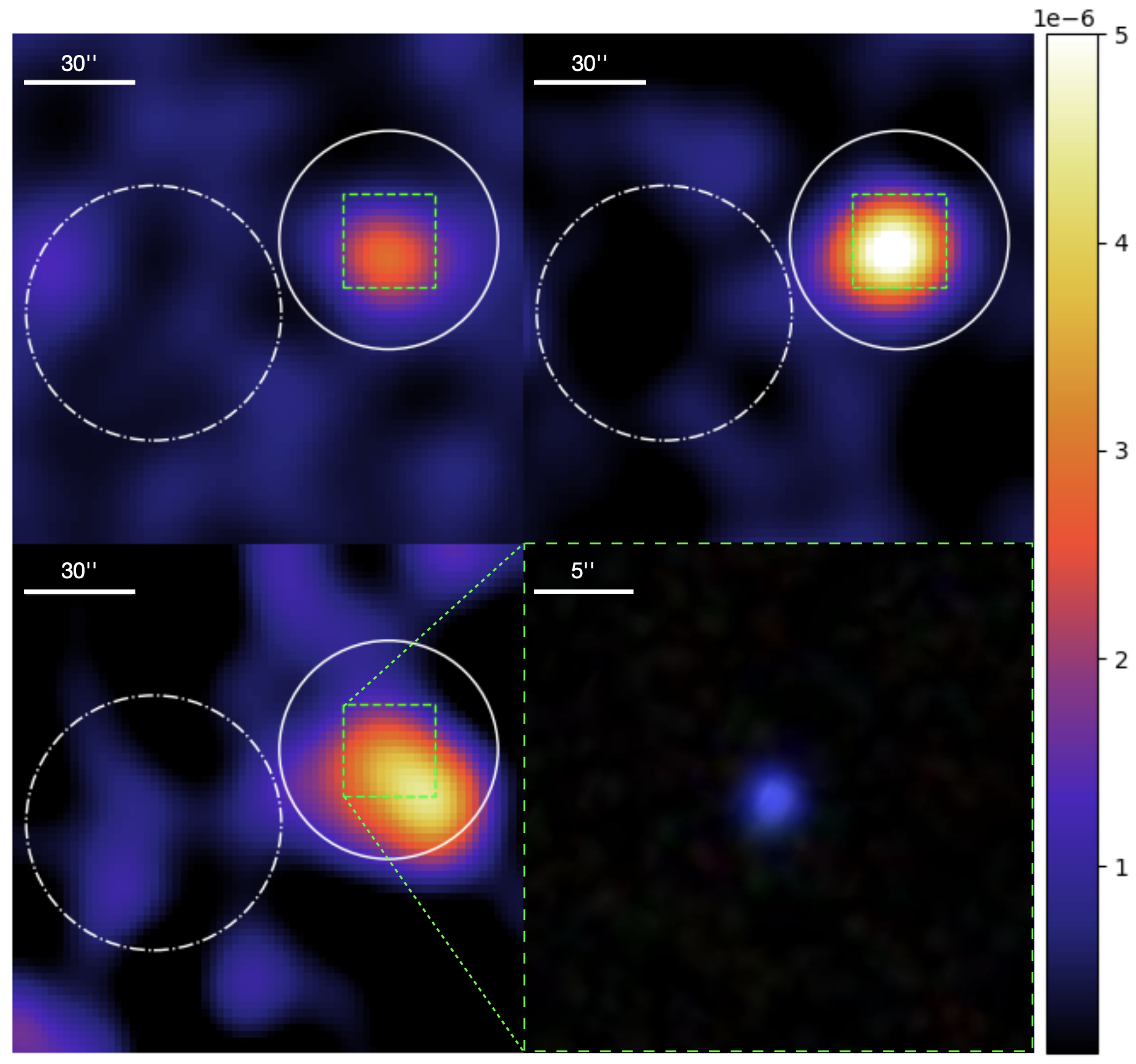}
   \end{tabular}
    \caption{X-ray vignetting-corrected background-subtracted images for BB1 (left) and BB8 (right) in the three used energy bands, 0.5-1 keV, 1–2 keV, 2–10 keV (from top left to bottom left), and the SDSS optical images (bottom right). The source extraction regions (solid white open circles) and background regions (dot-dashed white open circles) are denoted in the figures (details of the extraction regions can be found in the Appendix). The dashed green open rectangles, present in each X-ray image, indicate the regions corresponding to the SDSS images. The X-ray images are given in linear scale using the minima and maxima of the image cutouts. The colour scale denotes the pixel intensity, i.e. the scaled and weighted count rate per pixel. The SDSS images are shown at a different size scale compared to the X-ray images since a larger field of view would not provide adequate detail of the galaxies.}
              \label{fig:ds9blueberry}%
    \end{minipage}
    \end{figure*}

We analysed the X-ray spectra of the X-ray detected sources with the XSPEC spectral fitting software \citep[v12.12,][]{Arnaud1996}, using the energy range of $0.5-10$ keV. When a background model is not included in the spectral fitting, XSPEC is using W-statistics \citep{Wachter1979}, which is also referred to as modified \textit{C}-statistics\footnote{See the XSPEC manual by \citealt{Arnaud2022}.}
For the fitting, the combined EPIC spectra were binned using the \texttt{ftgrouppha} tool to contain at least 1 count in a single bin, which is recommended for the XSPEC implementation of the W-statistics. An absorbed power law model, \texttt{phabs*powerlaw}, was used for the spectral model, $I_{\nu} \approx \nu ^{-\alpha}$ with frequency $\nu$ and spectral slope $\alpha$. The abundances were set to values from \citet{Anders1989} and the \citep{Verner1996} photo-ionisation cross-sections were used. In X-ray astronomy, the photon index ($\Gamma$; where
$\Gamma = \alpha + 1$) is used to describe the power-law slope and it is associated with the X-ray emission produced by typical X-ray binary (XRB) populations. For the XSPEC analysis, $\Gamma$ and the normalisation factor were left to vary, and the uncertainties on the fitted parameters were calculated using the \texttt{err} command in the 90$\%$ confidence ranges for each parameter. The hydrogen column density $\mathrm{N}_\mathrm{H}$ was set to values calculated from the galaxy coordinates via the nH calculator\footnote{https://heasarc.gsfc.nasa.gov/cgi-bin/Tools/w3nh/w3nh.pl} since we assume the absorption to be only the neutral absorption in our Galaxy. The fluxes and luminosities were then determined using the \texttt{flux} and \texttt{lum} commands in XSPEC.

For the rest of the sources with low count rates and thus undetected by the SAS detection script, Bayesian analysis in low count regimes had to be applied, mainly because we cannot directly subtract the background from the source spectra. We make use of the Bayesian Estimation of Hardness Ratios (\texttt{BEHR}\footnote{http://hea-www.harvard.edu/astrostat/BEHR/index.html}) code \citep{Park2006}
to determine the posterior probability distribution of the counts in each of the undetected sources.

The same process of defining the source and background regions as for the detected sources was applied (see above). We made use of the \texttt{SAOImage DS9} \citep{ds9} \texttt{region statistics} tool on the images (energy range being 0.5-10 keV) and measured the number of counts and the area of the previously defined regions. As no prior information about the sources was known, we have used the non-informative Jeffrey’s prior distribution ($\Phi =1/2 $). The X-ray flux for each source was determined via the use of the \texttt{WebPIMMS}\footnote{https://heasarc.gsfc.nasa.gov/cgi-bin/Tools/w3pimms/w3pimms.pl} tool. To convert the count rate posteriors to fluxes, a power law model was used. The photon index was assumed to be $\Gamma = 1.9$, as it is the usual value for star-forming galaxies (see \citealp{Basu-Zych2013b}). The X-ray luminosity was then calculated as $L_{\mathrm{X}}= 4\pi D_{L}^2F_x$, where $D_{L}$ is the luminosity distance and $F_x$ is the X-ray flux measured in $0.5-8$\,keV.
The $\Lambda$CDM model is assumed throughout this work: Hubble constant $H_0=(67.4\:\pm \:0.5)$ $\mathrm{km\:s}^{-1}\:\mathrm{Mpc}^{-1}$ and matter density parameter $\Omega_m=0.315$ \citep{PlanckCollab2020}.

\begin{figure}
   \centering
   \includegraphics[width=9cm]{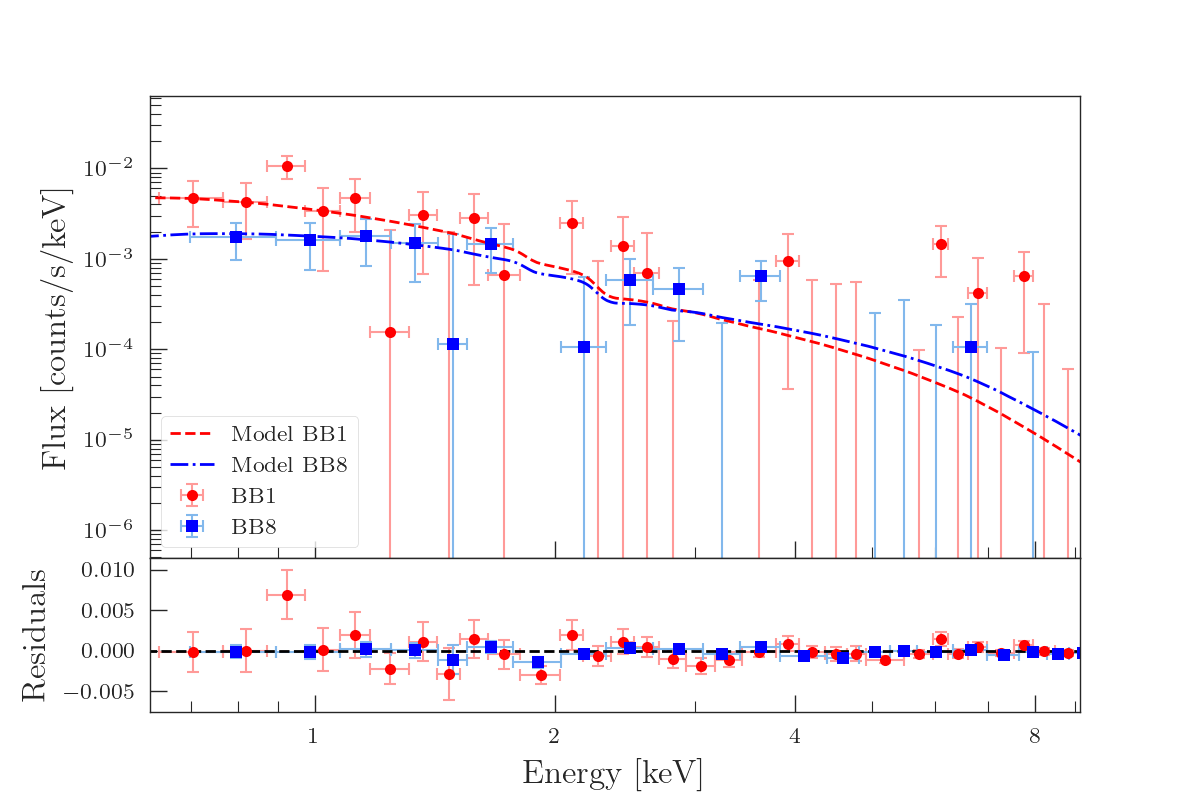}
   \caption{X-ray {\xmm} EPIC combined folded spectra of BB1 (filled red circles) and BB8 (filled blue squares) overplotted with their respective best-fit models (absorbed power-law) in red dashed for BB1 and blue dot-dashed for BB8. The residuals after the model subtraction are plotted in the bottom panel. The plotted data is binned to 20 counts per bin using the \texttt{setplot} \texttt{rebin} tool in XSPEC for plotting purposes only (the fitting was done with data only binned to have at least 1 count per bin).
   }
              \label{fig:Spectra}%
    \end{figure}

\section{Results}
\label{Results}

The measured X-ray fluxes and derived X-ray luminosities for detected sources (using XSPEC) as well as the measured upper limits for the undetected ones (determined using the \texttt{BEHR} code and using a 68\% confidence interval (CI) for the upper limits) are summarised in Table \ref{tab:results}. The last column lists the predicted values of X-ray luminosities from the \citet{Brorby2016} relation.
Only two sources, BB1 and BB8, have a significant detection revealed by the aforementioned SAS detection script. 
They both have the X-ray luminosity $\log\,L_{\mathrm{X}} \gtrsim 40.4$ erg s$^{-1}$, but the X-ray luminosity for BB8 is about five times larger than the expected one from SFR and metallicity. Both sources are detected in the softest X-ray bands (0.5-1 and 1-2 keV), and only BB8 is detected also in the harder 2-10 keV band. The results of the XSPEC spectral analysis using the absorbed power-law model are given in Table \ref{tab:fitresults}. We obtained good fits for both BBs, with the $\Gamma \approx 1.9$ for BB1 and $\Gamma \approx 1.7$ for BB8. The X-ray spectra of our two BBs are shown in Fig.~\ref{fig:Spectra} along with their best-fit power-law models and their residuals. We have also tested an APEC spectral model, but we did not obtain any better fit. 

\begin{figure*}
   \centering
   \includegraphics[width=9cm]{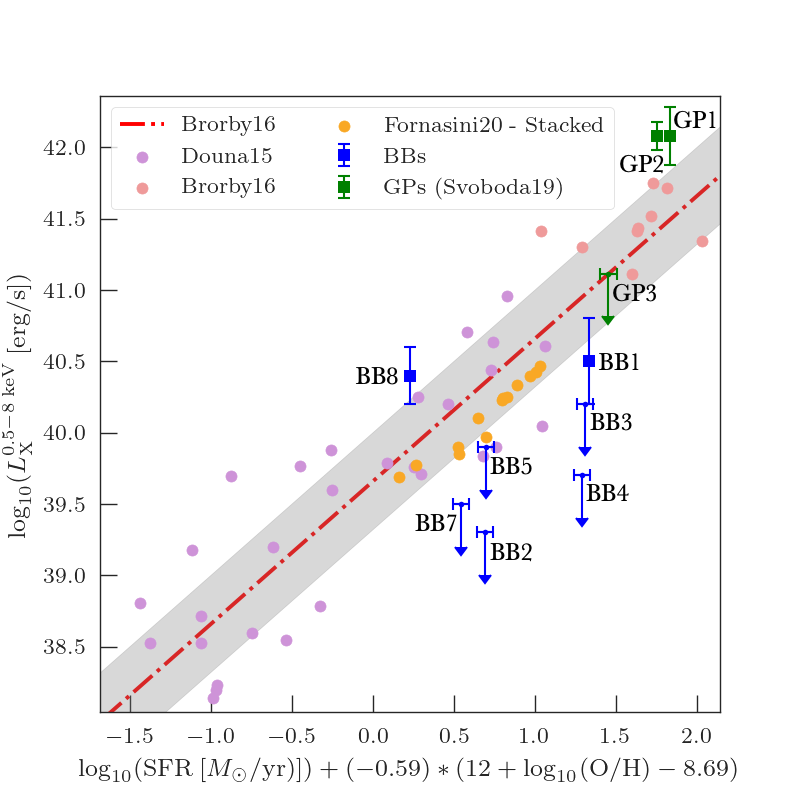}
   \includegraphics[width=9cm]{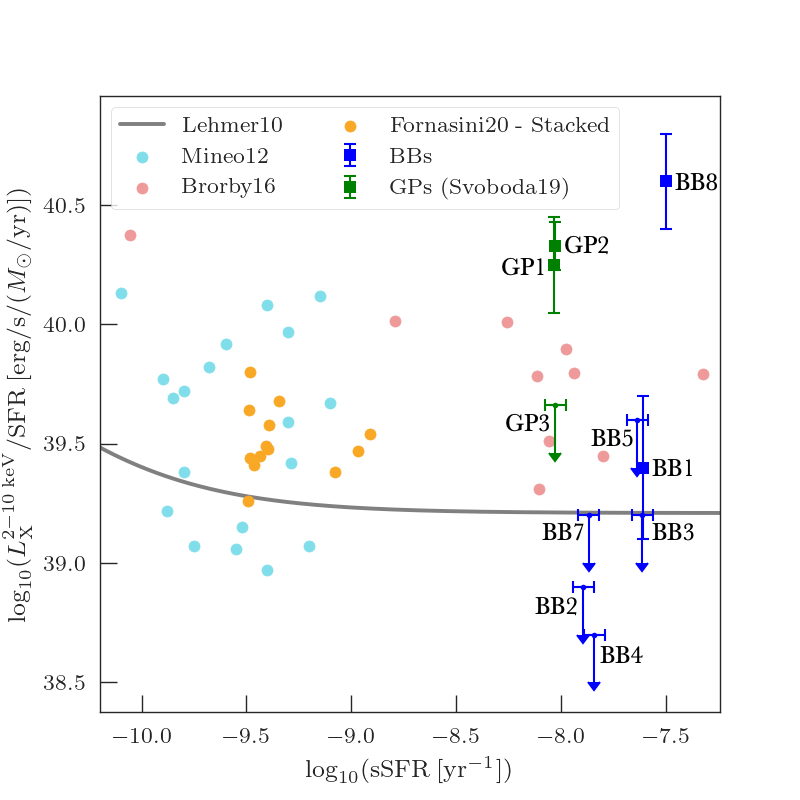}
   \caption{\textit{Left:} X-ray luminosity $L_{\mathrm{X}}$ as a function of SFR and metallicity for the BBs (filled blue squares and down-arrows) studied by {\xmm}. The empirical relation by \citet{Brorby2016} is shown as a red dot-dashed line and its 1{$\sigma$} deviation as a grey region. \textit{Right:} Our BB sample in the diagram of the X-ray luminosity over the SFR as dependent on the sSFR. The solid horizontal line represents the \citet{Lehmer2010} relation. The GPs sample by \citet{Svoboda2019} (filled green squares and down-arrows) is in both diagrams plotted for comparison along with few other samples of \citet{Mineo2012} (filled turquoise circles only in the right plot), \citet{Douna2015} (filled purple circles only in the left plot), \citet{Brorby2016} (filled pink circles), and \citet{fornasini20} (filled dark yellow circles).
   }
              \label{fig:brorby}%
    \end{figure*}

For the two detected sources, we created the X-ray vignetting-corrected background-subtracted images for three energy bands, 0.5-1 keV, 1–2 keV, 2–10 keV, and we obtained the optical images from the SDSS DR16 SkyServer\footnote{https://skyserver.sdss.org/dr16/en/home.aspx} (Fig.~\ref{fig:ds9blueberry}). The BB1 has a strong detection in the 0.5-1 and 1-2 keV bands, while BB8 also shows a significant detection in the harder 2-10 keV band. The optical images reveal that BB1 has extended features, whereas BB8 appears to be very compact. Additionally, there is a source nearby to BB1, classified as a star in the SDSS DR16 SkyServer, and thus not contributing to the X-ray energy band.

The detection of BB1 and BB8 was also confirmed using the Bayesian analysis via the \texttt{BEHR} code. The \texttt{BEHR} code gave the same resulting fluxes and luminosities as the XSPEC analysis, the results of which were used in the rest of the paper. For the rest of the observed sources, only the Bayesian upper limits could be measured.

The BB X-ray measurements along with the empirical relation by \citet{Brorby2016} for star-forming galaxies are plotted in Fig.~\ref{fig:brorby} (left). Except for BB8, which is above, most of our sources are below the empirical relation, with BB2 and BB4 having their upper limits about an order of magnitude below the relation. For comparison, we also plot the GPs by \citet{Svoboda2019}, for which two sources were detected above the empirical relation and one measured upper limit shows as consistent. The right panel of Fig.~\ref{fig:brorby} shows $L_{\mathrm{2-10\,keV}}$/SFR vs. sSFR. While BB8 is above the scaling relation by \citet{Lehmer2010}, most of the other BBs are consistent or below. The two detected GPs are again significantly above the empirical relation. We also plot a few previously studied star-forming galaxy samples in Fig.~\ref{fig:brorby}.

\begin{figure}
   \includegraphics[width=9cm]{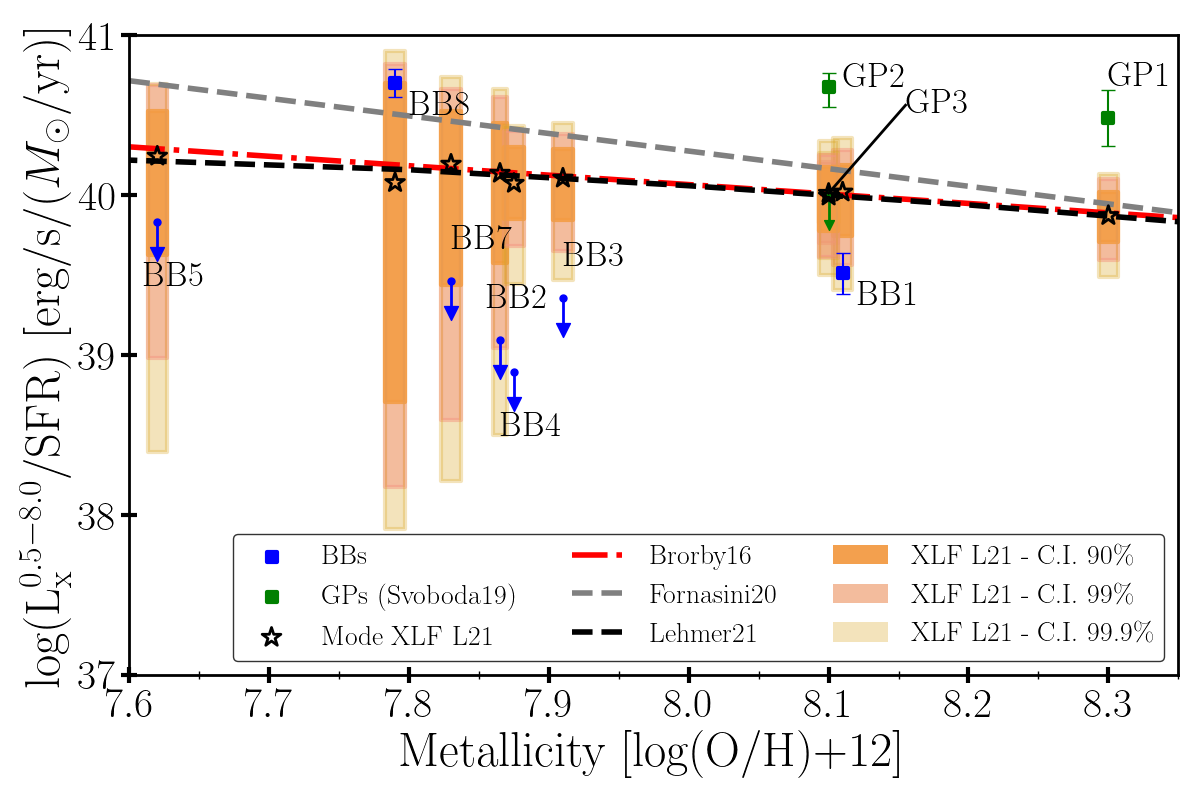}
   \caption{Distribution of the total expected X-ray luminosity per SFR due to the stochastic sampling of the HMXBs XLF from L21, as a function of the metallicity of each galaxy in our sample. Orange, pink and goldenrod shaded regions indicate the 90$\%$, the 99$\%$  and the 99.9$\%$ CI, respectively, of the expected $L_{\mathrm{X}}$/SFR distribution of each BB. Filled blue squares and down-arrows indicate the BBs galaxy sample of this work. We also plot with filled green squares and down-arrows the GPs sample from \citealp{Svoboda2019}. For comparison, we overplot with red dot-dashed line the scaling relation from \citealp[]{Brorby2016}, with gray dashed line the relation from \citealp[]{fornasini20}, and with black dashed line the relation from \citealp[]{lehmer21}.}
              \label{fig:Lx_SFR_Metallicity_stochasticity}%
    \end{figure}

A possible source for the large scatter of our BBs galaxy sample may be the stochastic sampling of individual XRBs associated with the galaxies. To investigate this assumption, we performed a simulation study following a similar approach as in \citet{anastasopoulou19}, \citet{Kouroumpatzakis2021}, and \citet{Kyritsis2024}. In particular, using the SFR and the gas-phase metallicity of each galaxy (Table \ref{tab:sample}) and by integrating their corresponding metallicity-dependent XLF from \citealp[]{lehmer21} (L21), we calculated the expected number of HMXBs for each galaxy as follows:
\begin{equation}
N_{exp} = \int^{L_{\rm{max}}}_{L_{\rm{min}}}\frac{dN_{HMXB}}{dL}\cdot dL .
\end{equation}

For the integration of the L21 XLF we assumed their best-fit parameters (from their Table 2), and the integration limits spanned between the $L_{\rm{min}} = 10^{36}\,\rm{erg}\,\rm{s^{-1}}$ and $L_{\rm{max}} = 5\times10^{41}\,\rm{erg}\,\rm{s^{-1}}$, representing the minimum and maximum luminosity, respectively. 
Subsequently, by sampling from a Poisson distribution with a mean equal to the number of expected HMXBs calculated above, we generated 20000 draws of the expected number of HMXBs ($N_{\rm{exp},i}^{\rm{inst}}$). Then, we sampled the HMXB XLF of each galaxy, by drawing the corresponding X-ray luminosities of $N_{\rm{exp},I}^{\rm{inst}}$ sources each time. The total X-ray luminosity for each of the 20000 simulations was then computed by summing the luminosity of each source drawn from the XLF for each instance. 

This analysis allowed us to simulate the X-ray emission of 20000 galaxies, considering the typical properties (SFR, metallicity) of the galaxies in our sample. At the same time, we accounted for fluctuations in the number of sources within each galaxy and stochastic effects associated with the sampling of their XLF. For each of these distributions, we calculated the mode, and the $90\%$, $99\%$, and the $99.9\%$ upper and lower confidence intervals. The probability of the X-ray luminosity of a given galaxy being greater or smaller than the observed value is given in the Appendix in Table \ref{tab:probabilities}, the probability distributions based on the stochastic sampling are plotted in Fig. \ref{fig:Lx_stochasticity_distributions} in Appendix.

In Fig. \ref{fig:Lx_SFR_Metallicity_stochasticity} we present the 90$\%$, the 99$\%$, and the 99.9$\%$ CIs of the total expected X-ray luminosity per SFR distribution due to the stochastic sampling of the HMXBs XLF from L21, as a function of the metallicity of each galaxy in our sample. We also show in blue the BBs sample of this work while with green the GPs sample from \citealp{Svoboda2019}. For comparison, we overplot the scaling relations from \citealp[]{Brorby2016}, \citealp[]{fornasini20}, and \citealp[]{lehmer21}. Only one of the BBs, BB8, is above the empirical relation by \citet{lehmer21}. More specifically, the probability of its luminosity being an outcome of stochastic behaviour is less than 1.2\%. On the contrary, the rest of the BBs are significantly below the relation. The BB5's upper limit is within the 90$\%$ CI from the expected value, but all the other upper limits are beyond. The detected BB1 and the upper limits of BB2 and BB7 lie in the 99$\%$ CI region. Two sources, BB3 and BB4, have their upper limits even beyond the lower bound of 99.9$\%$ and thus cannot be attributed to stochasticity effects at all. The jump in the distribution between BB2 and BB4 is due to them having the same metallicity, but different SFR. The two bright GPs both lie beyond the upper bounds of the stochasticity sampling. The upper limit of the weakest GP, on the other hand, shows as consistent with the \citet{lehmer21} relation.

\section{Discussion}
\label{Discussion}

In this paper, we have studied the first X-ray observations of the extremely low-metallicity compact and highly star-forming young galaxies, the BBs. The X-ray data was obtained with the {\xmm} satellite. The planned exposures were calculated to be sufficient for the detection of the expected X-ray flux based on empirical relations between X-ray flux, star formation, and metallicity. However, only 2 out of 7 sources have been detected (with 1 non-detection probably due to high background flare contamination). For the two detected sources, the X-ray flux and luminosity could be properly measured and the spectral fitting applied. 

The luminosity of these two detected sources is about ${\log}L_{\mathrm{X}}\approx 40.5$ erg s$^{-1}$, with one (BB1) being close to the expected luminosity estimated from empirical relations for star-forming galaxies, and the other (BB8) showing an X-ray excess five times the expected value. To explore the possibility of contamination from a background AGN or source confusion, we adopted a method similar to \citet{Svoboda2019}. We estimated the number of background sources expected to have the same flux as the two BBs, which corresponds to the detection probability of such background AGN. The BB fluxes in the 0.5-2\,keV band are log$F_{\mathrm{X}} \approx - 14.5$ erg s$^{-1}$ cm$^{-2}$ for BB1 and log$F_{\mathrm{X}} \approx - 14.6$ erg s$^{-1}$ cm$^{-2}$ for BB8, which can then be compared to the values in Table 3 in \citet{Mateos2008}. Given the higher probability of finding a fainter source, we use the values for a flux of log$F_{\mathrm{X}} \approx - 14.7$ erg s$^{-1}$ cm$^{-2}$. For an area of $1\,\mathrm{deg}^2$, the probability of finding a source with this flux is $N(>S)\approx474$ \citep{Mateos2008}. For our source extraction regions of 30 arcsecs (0.00022 $\mathrm{deg}^2$) it translates to $N(>S)\approx0.1$. Therefore, the probability of finding a random background AGN in our source regions is quite low. The only nearby source is the star 13 arcsec south of BB1 (as classified by the SDSS DR16 SkyServer), but it is not expected to contribute to the X-ray flux at all.

Two sources, BB1 and BB2, were previously measured by a targeted Chandra observation \citep{Wang2016}. The BB1 was detected with a measured flux by \citet{Wang2016} and its luminosity have been constrained by \citet{Latimer2021} to be ${\log}L_{\mathrm{X}}\approx 40.5$ erg s$^{-1}$ in 0.5-2\,keV band, and ${\log}L_{\mathrm{X}}\approx 40.8$ erg s$^{-1}$ in 2-10\,keV band. For the rest of the BBs, which were undetected by the standard methods, only the Bayesian-based upper limits on X-ray flux and luminosity could have been obtained. The upper limits of the luminosity of these sources are typically less than 10$^{40}$ erg s$^{-1}$.

The fact that our sample of BBs is mostly X-ray underluminous compared to the empirical relations using their SFR, stellar mass, and metallicity values, is a somewhat surprising result. The relation by \citet{Brorby2016} was built for galaxies dominated by HMXBs, that is the young star-forming galaxies. The sSFRs of BBs range from $-7.9$ to $-7.5$ (See Fig.~\ref{fig:brorby}), which is extremely high, signifying the BBs are very young and star-forming galaxies, and puts them far into the HMXB dominated region (sSFRs $> -10$). 
Following the previous study of GPs \citep{Svoboda2019}, which found a significant X-ray excess in two out of three sources, an enhanced X-ray luminosity of BBs in more than one source could have been expected, especially, since the BBs are more extreme than GPs. However, our results suggest the opposite, the BBs do not follow the empirical laws for local star-forming galaxies and could have lower X-ray luminosity on average. This result indicates that the higher emission of BB8 and the two GPs is likely of different origin than from pure star-formation activity.

For the two GPs, \citet{Svoboda2019} proposed that the enhanced X-ray luminosity can be attributed to a hidden AGN. The X-ray emission was too large to be produced by a standard XRB population. Both a larger population of X-ray binaries or ULXs were considered but concluded as unlikely since the GPs have the X-ray luminosity of the order of $\sim$10$^{42}$ erg s$^{-1}$ \citep{Svoboda2019}, which is too high even for ULXs with their typical X-ray luminosity of the order of $\sim$10$^{39}$ erg s$^{-1}$ (e.g. \citealt{Kaaret2017}). For BB8, the X-ray luminosity $\sim$10$^{40}$ erg s$^{-1}$ is not as high, and the enhanced luminosity could be due to an additional X-ray source (several or bright ULXs, or a larger population of HMXBs). However, we note that the stochasticity simulations do include contributions from ULXs, and therefore only the rare extreme ULXs would be a possibility for high X-ray emission explanation.

We have compared the measured luminosity and the upper limits of observed BBs to different empirical relations by \citet{Lehmer2010}, \citet{Brorby2016}, \citet{fornasini20} and \citet{lehmer21}, but with similar conclusions. Our BB8 shows enhanced X-ray luminosity with respect to all considered relations and is inconsistent at almost 99$\%$ confidence level with being explained due to stochastic sampling of the HMXBs XLF. The X-ray luminosity of BB1 is either underluminous or consistent with them, and the upper limits of the rest BB2$-$BB7 are mostly below the empirical curves (see Figs.~\ref{fig:brorby} and \ref{fig:Lx_SFR_Metallicity_stochasticity}). The deviation is seemingly less pronounced for the relation by \citet{Lehmer2010}, shown in the right panel of Fig.~\ref{fig:brorby}. However, this is likely just coincident since the other samples are more appropriate for the comparison with BBs that are largely in the HMXB-dominated regime ($\log({\textrm{sSFR}}) > -10$). The relation by \citet{Lehmer2010} considered both LMXBs and HMXBs, while \citet{Brorby2016} and \citet{lehmer21} focused their sample only on the galaxies whose X-ray flux is dominated by HMXBs and they took into account also the effects of metallicity (though in \citealp{Lehmer2010}, the stellar mass is taken into account and thus the metallicity is indirectly considered through the mass-metallicity relation by \citealp{Tremonti2004}). 
The lowest-metallicity source BB5 was too underexposed to have a robust detection due to high background flares. Nevertheless, this source is illustrative of how significant the metallicity effect is in the considered empirical relations. Compared to \citet{Lehmer2010}, the upper limit is consistent with the expected X-ray luminosity. However, it is below the empirical relations that took the metallicity directly into account.

The extremely high sSFRs of BBs indicate that their dominating stellar population is very young, much younger than the star-forming galaxies used for the empirical relations. This leads to a hypothesis that our BBs may not be old enough to give rise to a significant population of X-ray binaries. This effect was studied by \citet{Shtykovskiy2007,Antoniou2010,Antoniou2019} for Be-XRBs in the Magellanic Clouds. They showed that for a population of 10 Myr, there are an order of magnitude fewer HMXBs than in a population of 40 Myr. However, \citet{Fragos2013} showed, that for XRB populations younger than 10 Myr, the X-ray luminosity seems to be higher as the HMXBs tend to be more luminous. Therefore, our BBs would have to be even younger (probably less than 5 Myr) for this to have a significant effect. As \citet{Linden2010} showed, the characteristics and star formation history of young populations of bright HMXBs depend strongly on metallicity and formation channels. 
Thus, this could also be an effect in our galaxy sample as our metallicities are rather low (log(O/H)+12 < 8.1). Although we do not know the exact star-formation history or stellar population ages of our studied sources, if the high sSFR of BBs indicates that their stellar populations are extremely young, this might be a reason for the deficit of the X-ray luminosity especially for BB4, which is significantly below the empirical relations, even after taking the stochasticity into account (see Fig.~\ref{fig:Lx_SFR_Metallicity_stochasticity}). A detailed analysis of the optical spectra of these galaxies could provide insights into their star formation histories and help explain the observed low X-ray luminosities.

The obtained results might also be relevant for the high-redshift galaxies at the Epoch of Reionisation. The BBs with their extreme sSFR due to high SFR and compactness, can be considered reminiscent of the first early-Universe galaxies similarly to GPs whose spectra are noticeably similar to the early high-redshift galaxies, recently observed by JWST \citep{Schaerer2022,Rhoads2023}. The higher X-ray flux of BB8, especially when compared to the other X-ray underluminous BBs, might suggest an alternative explanation: the elevated X-ray flux could be due to AGN activity, as was proposed for two GPs. But only for the GPs, the stochasticity effects cannot explain the elevation at more than 99.9$\%$ confidence level (see Fig.~\ref{fig:Lx_SFR_Metallicity_stochasticity}), and the AGN scenario thus remains as one of the most viable interpretations for these two GPs.

The presence of AGN in dwarf galaxies has been revealed in several other sources \citep{Greene2004, Reines2013, Mezcua2018, Mezcua2020, Birchall2020, Reines2022, Mezcua2023}. Nonetheless, AGNs in dwarf galaxies are difficult to detect as they are expected to be intrinsically low luminosity. Their X-ray luminosities are expected to be lower than $\sim$10$^{40}$ or $\sim$10$^{41}$ erg s$^{-1}$), with a significant contribution from the thermal emission attributed to star formation that contaminates the signal on lower energies. The AGN presence in low-metallicity compact galaxies with extremely high sSFR, such as GPs and BBs, is very rare and can improve our knowledge of the first galaxies in the Universe, their evolution and to properly understand the crucial source of the ionising radiation during the cosmic reionisation. A more detailed investigation of potential AGN candidates among them, as well as a systematic analysis of a larger sample of these local analogues, are needed to shed light on the power of the galaxies in the early Universe. One such method might include the optical variability of these sources, as studied by \citeauthor{Baldassare2018} (\citeyear{Baldassare2018}, \citeyear{Baldassare2020}).

\section{Conclusions}
\label{Conclusions}

We have studied 7 BBs in X-rays using data from the {\xmm} satellite. Only 2 sources have been detected and their X-ray luminosity could be obtained. The luminosity was constrained to be ${\log}L_{\mathrm{X}}=40.5\,\pm\,0.3$ erg s$^{-1}$ for BB1 and ${\log}L_{\mathrm{X}}=40.4\,\pm\,0.2$ erg s$^{-1}$ for BB8, respectively. For the remaining sources, only the Bayesian-based upper limits were measured, one source (BB3) has an X-ray luminosity upper limit of ${\log}L_{\mathrm{X}}\textless{} 40.2$ erg s$^{-1}$, the rest are below ${\log}L_{\mathrm{X}}\sim{}39.9$ erg s$^{-1}$. Comparison with different empirical relations showed that except for the brightest source (BB8), most of the BBs are below the X-ray luminosities predicted by empirical relations. By applying the stochasticity analysis, we showed that the brightest source (BB8) is at a 99$\%$ confidence level inconsistent with the expectations, which could hint at the presence of an additional X-ray source, such as an AGN or an extreme ULX. Oppositely, the second detected source (BB1) is at more than 99$\%$ inconsistent with the lower bounds of the stochasticity sampling. Two sources (BB3 and BB4) are completely out of the lower bounds and therefore cannot be explained by the stochasticity with more than 99.9\% probability. Therefore, our results indicate that the BBs have either much larger scatter and/or are markedly X-ray underluminous. The insufficient X-ray luminosity might be due to the BBs not having had enough time to develop a significant XRB population. A larger sample of BBs, combined with their optical spectra and variability measurements, would enhance our understanding of these sources.

\begin{acknowledgements}
      We sincerely thank the anonymous referee for their insightful comments, which have significantly improved the quality of the manuscript. This work was supported by the Czech Science Foundation project No.22-22643S. MC and POP were partly supported by the High Energy National Programme (PNHE) of the French National Center of Scientific Research (CNRS) and the French Spatial Agency (CNES).

Software used: 

\texttt{Numpy} \citep{Numpy}, \texttt{matplotlib} \citep{Matplotlib}, \texttt{astropy} \citep{Astropy2013,Astropy2018,Astropy2022}
      
\end{acknowledgements}

\bibliographystyle{aa}
\bibliography{refs}

\begin{appendix}
\section{Details of source and background regions}
The details of the source and different background extraction regions for all seven BBs are summarised in Table \ref{tab:regions}.

\numberwithin{table}{section}
\setcounter{table}{0}
\renewcommand{\thetable}{A\arabic{table}}
\begin{table}[h!]
\centering
\caption{Details of the source and background extraction regions}
\begin{tabular}{cllllll}
\hline
\hline
\rule{0cm}{0.4cm}
{Source} & \multicolumn{3}{l}{Source extraction region} &  \multicolumn{3}{l}{Background extraction region} \\
             & RA                 & Dec                & Rad      & RA                 & Dec                & Rad      \\
& [h:m:s]                 & [d:m:s]               & [arcsec]       & [h:m:s]               & [d:m:s]                 & [arcsec]       \\
\hline
\rule{0cm}{0.4cm}
BB1          & 17:35:01.228     & +57:03:08.460    & 30     &  17:34:56.360       & +57:04:02.882      & 35       \\
~BB2           & 15:09:34.173     & +37:31:46.128    & 30     & 15:09:29.509       & +37:32:21.106      & 33       \\
~BB3          & 2:40:52.195     & -8:28:27.480     & 30     &  2:40:56.328        & -8:28:09.511       & 30       \\
~BB4           & 8:51:15.650      & +58:40:55.020    & 30     &  8:51:22.408        & +58:40:16.479      & 32       \\
~BB5           & 1:46:53.306     & +3:19:22.368     & 30     &  1:46:57.057        & +3:19:46.630       & 30       \\
~BB7           & 15:56:24.475     & +48:06:45.792    & 30     &  15:56:17.777       & +48:06:46.089      & 32       \\
~BB8           & 8:25:40.449      & +18:46:17.220    & 30     &  8:25:45.070        & +18:45:57.170      & 35  \\
\hline
\end{tabular}
\begin{tablenotes}
    \item {\textbf{Notes:}} The coordinates are given for the centres of the regions in the FK5 system (J2000).
\end{tablenotes}
\label{tab:regions}
\end{table}
\FloatBarrier

\section{Stochasticity probabilities and X-ray luminosity distributions}
The results of the probability analysis of the stochasticity sampling are summarised in Table \ref{tab:probabilities}, where the probability of stochastically observing a galaxy, having the same properties as the observed one, with X-ray luminosity greater or smaller than the observed value is given. In addition, in Fig. \ref{fig:Lx_stochasticity_distributions} we plot the histograms of the X-ray luminosity distribution over SFR based on the stochasticity simulations. 

\numberwithin{table}{section}
\setcounter{table}{1}
\renewcommand{\thetable}{B\arabic{table}}
\begin{table}[h!]
\centering
\caption{Probability for a galaxy, with the same physical properties as the observed one, having 
X-ray luminosity greater or less than the observed value (i.e. Lx\textgreater{}Lobs or Lx\textless{}Lobs) due to stochastic sampling}
\begin{tabular}{lcccccccccc}
\hline
\hline
\rule{0cm}{0.4cm}
                     & BB1   & BB2   & BB3 & BB4 & BB5   & BB7   & BB8   & GP1 & GP2 & GP3   \\
\hline
\rule{0cm}{0.3cm}
Lx\textgreater{}Lobs & 0.997 & 0.989 & 1.0 & 1.0 & 0.808 & 0.912 & 0.013 & 0.0 & 0.0 & 0.408 \\
~Lx\textless{}Lobs    & 0.003 & 0.011 & 0.0 & 0.0 & 0.192 & 0.088 & 0.987 & 1.0 & 1.0 & 0.592 \\
\hline
\end{tabular}
\label{tab:probabilities}
\end{table}
\FloatBarrier

\numberwithin{figure}{section}
\setcounter{figure}{0}
\renewcommand{\thefigure}{B\arabic{figure}}
\begin{figure*}[h!]
   \centering
   \includegraphics[width=1\textwidth]{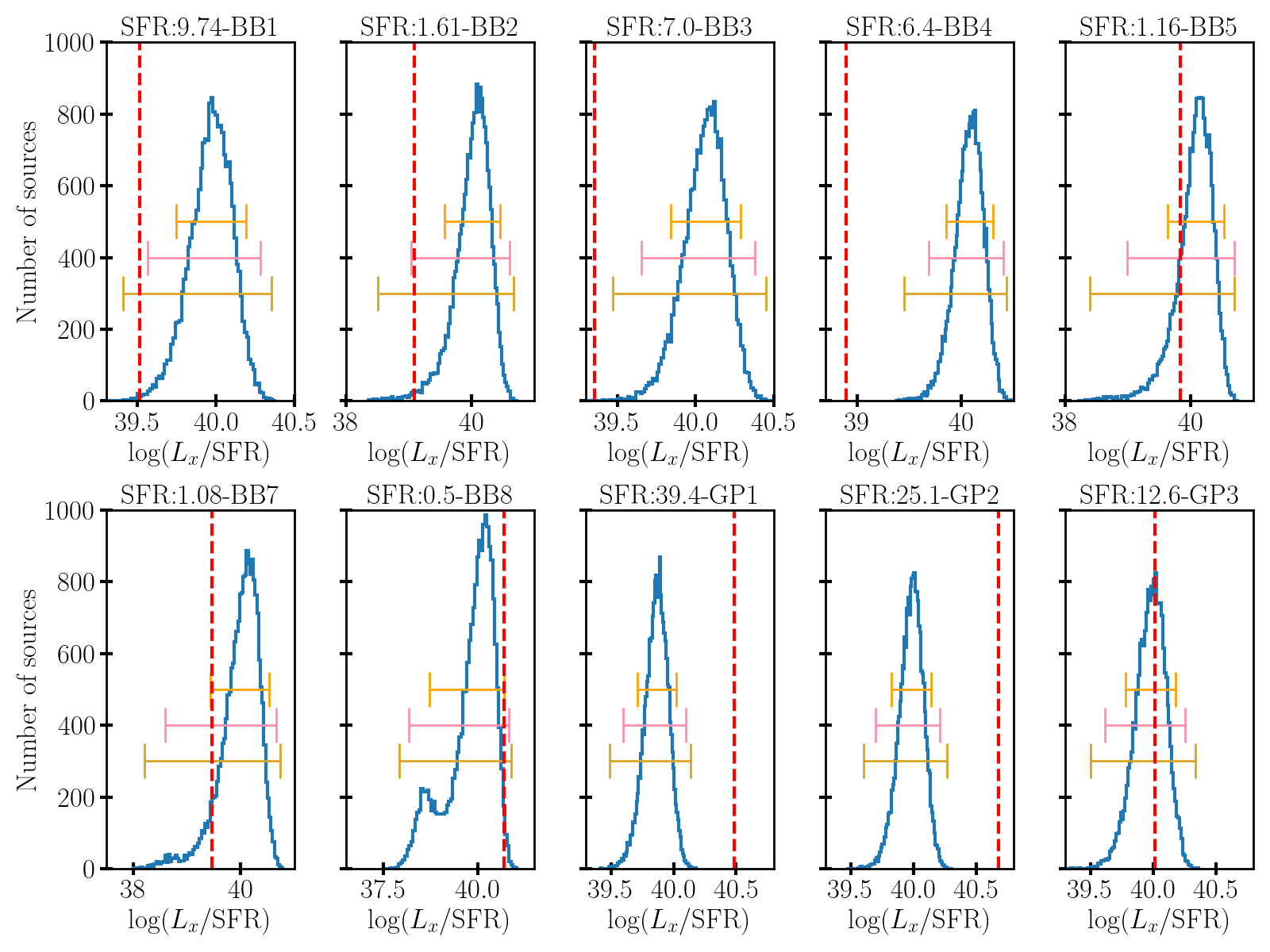}
   \caption{Histograms of the X-ray luminosity distribution over SFR based on the stochasticity simulations for 7 BBs and 3 GPs. The red dashed line indicates the observed X-ray luminosity. The orange, pink and goldenrod bars indicate the lower and upper bounds of the CI 90\%, 99\%, and 99.9\%.
   }
              \label{fig:Lx_stochasticity_distributions}
    \end{figure*}
\FloatBarrier

\end{appendix}
\end{document}